\documentclass[aps,pra,showpacs,twocolumn]{revtex4-1}
\usepackage{amssymb}
\usepackage{amsmath}
\usepackage{graphicx}
\usepackage{epsfig}

\begin{document}

\title{Chern number in Ising models with spatially modulated real and
complex fields}
\author{C. Li${}^1$, G. Zhang${}^2$ and Z. Song${}^1$}
\email{songtc@nankai.edu.cn}
\affiliation{${}^1$School of Physics, Nankai University, Tianjin 300071, China \\
${}^2$College of Physics and Materials Science, Tianjin Normal University,
Tianjin 300387, China}

\begin{abstract}
We study an one-dimensional transverse field Ising model with additional
periodically modulated real and complex fields. It is shown that both models
can be mapped on a pseudo spin system in the $k$ space in the aid of an
extended Bogoliubov transformation. This allows us to introduce the
geometric quantity, the Chern number, to identify the nature of quantum
phases. Based on the exact solution, we find that the spatially modulated
real and complex fields rearrange the phase boundaries from that of the
ordinary Ising model, which can be characterized by the Chern numbers
defined in the context of Dirac and biorthonormal inner products,
respectively.
\end{abstract}

\maketitle

\section{Introduction}

Characterizing the quantum phase transitions (QPTs) is of central
significance to both condensed matter physics and quantum information
science. Exactly solvable quantum many-body models are benefit to
demonstrate the concept and characteristic of QPTs. Recently, topological
phases and phase transitions \cite{Wen} have attracted much attention in
various physical contexts. In general, QPTs are classified two types,
characterized by topologically nontrivial properties in the Hilbert space,
and by the local order parameters associated with symmetry breaking,
respectively. A topological state typically features a topological
invariant, the Chern number. In recent work \cite{ZG PRL}, it turns out that
the local order parameter and topological order parameter can coexist to
characterize the quantum phase transitions.

Our aim is extending the result on the connection between the quantum phase
diagram and geometric quantity to\ more generalized systems, which are
exactly solvable models of spin systems with a little complicated transverse
field. The Hamiltonians are Hermitian and non-Hermitian, depending on the
magnetic field affecting the whole system. In both cases, the conventional
QPT occurring at zero temperature, i.e., the groundstate energy density
experiences a divergence \cite{S.Sachdev,LC1 PRA}, rather than the
appearance of complex eigen energy \cite{Bender 02,Bender 98,Bender 99,ZXZ1
PRA,ZXZ2 PRA}.

In this paper, we investigate an one-dimensional transverse field Ising
model with additional periodically modulated real and complex fields. It is
shown that both models can be mapped on a pseudo spin system in the $k$
space with the aid of an extended Bogoliubov transformation. This allows us
to introduce the geometric quantity, the Chern number, to identify the
nature of quantum phases. Based on the exact solution, we find that the
spatially modulated real and complex fields rearrange the phase boundaries
from that of the ordinary Ising model, which can be characterized by the
Chern numbers defined in the context of Dirac and biorthonormal inner
products, respectively.

This paper is organized as follows. In section \ref{Hamiltonian and
solutions}, we present the models and the solutions. In section \ref{Chern
numbers}, we calculate the Chern numbers in different quantum phases.
Section \ref{Summary} summarizes the results and explores its implications.

\begin{figure}[tbp]
\includegraphics[ bb=51 520 559 805, width=0.45\textwidth, clip]{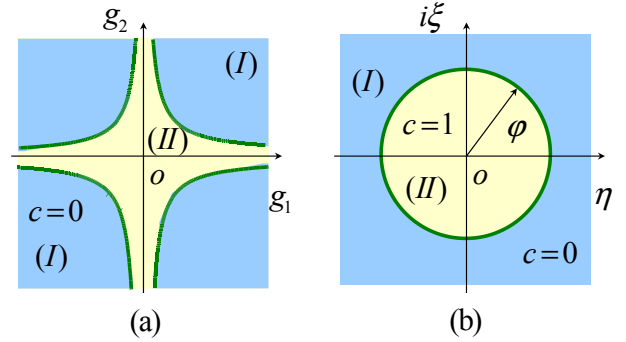}
\caption{(Color online) Phase diagrams of Hermitian and non-Hermitian Ising
models. (a) The staggered field strength $g_{1}$ and $g_{2}$ are real and
the phase boundary is $\left\vert g_{1}g_{2}\right\vert =1$. (b) The
staggered field strength is complex with real part $\protect\eta $ and
imaginary part $\pm \protect\xi $. The phase boundary is the circle $\protect\eta ^{2}+\protect\xi ^{2}=1$. In both cases, the Chern number can be
employed to characterized the quantum phases in the regions I and II.} \label%
{fig1}
\end{figure}

\section{Hamiltonian and solutions}

\label{Hamiltonian and solutions}

We start our investigation by considering a Ising ring with spatially
modulated real and complex fields%
\begin{equation}
H=-\underset{j=1}{\overset{2N}{\sum }}\left( \sigma _{j}^{z}\sigma
_{j+1}^{z}+g_{_{j}}\sigma _{j}^{x}\right) ,  \label{H_general}
\end{equation}%
where the external field $g_{_{j}}$\ can be either real or complex. For real
field, we take $g_{_{j}}=g_{_{1}}$\ ($g_{_{2}}$)\ for odd (even) $j$, while
complex field $g_{_{1}}=\eta -i\xi $\ and $g_{2}=\eta +i\xi $\ ($i=\sqrt{-1}$%
) with $\eta $\ and $\xi $\ being real numbers. Here $\sigma _{j}^{\lambda }$
($\lambda =x,$ $y,$ $z$) are the Pauli operators on site $j$, and satisfy
the periodic boundary condition $\sigma _{j}^{\lambda }\equiv \sigma
_{j+2N}^{\lambda }$.

The groundstate properties of this model in its Hermitian version have been
studied recently \cite{Giorgi1,Canosa}, while its non-Hermitian version is
investigated in Refs. \cite{Giorgi2,LC1 PRA,LC2 PRA}. Although the
transverse field is non-Hermitian, it turns out that the QPTs still have the
features in a Hermitian one, such as the divergence geometric phase \cite%
{LC1 PRA} and vanishing fidelity \cite{LC2 PRA} near the critical points.

In the following, we will diagonalize the Hermitian and non-Hermitian models
respectively. In both cases, one can perform the Jordan-Wigner
transformation \cite{P.Jordan}%
\begin{eqnarray}
\sigma _{j}^{+} &=&\prod\limits_{l<j}\left( 1-2c_{l}^{\dag }c_{l}\right)
c_{j}, \\
\sigma _{j}^{x} &=&1-2c_{j}^{\dag }c_{j}, \\
\sigma _{j}^{z} &=&-\prod\limits_{l<j}\left( 1-2c_{l}^{\dagger }c_{l}\right)
\left( c_{j}+c_{j}^{\dagger }\right) ,
\end{eqnarray}%
to replace the Pauli operators by the fermionic operators $c_{j}$. In this
paper, we focus on the system in the thermodynamic limit, in which the
difference between even and odd numbers of fermions can be neglected. The
Hamiltonian can be expressed as%
\begin{eqnarray}
H &=&-\sum\limits_{j=1}^{2N-1}(c_{j}^{\dag }c_{j+1}+c_{j}^{\dag
}c_{j+1}^{\dag })+c_{2N}^{\dag }c_{1}+c_{2N}^{\dag }c_{1}^{\dag }  \notag \\
&&+\mathrm{h.c.}-J\sum\limits_{j=1}^{2N-1}g_{j}(1-2c_{j}^{\dagger }c_{j}).
\label{H_fermi}
\end{eqnarray}%
The Fourier transformation of two sub-lattices is%
\begin{equation}
c_{j}=\frac{1}{\sqrt{N}}\sum\limits_{k}e^{ikj}\left\{
\begin{array}{cc}
e^{ik/2}\alpha _{k}, & \text{even }j \\
\beta _{k}, & \text{odd }j%
\end{array}%
\right. ,
\end{equation}%
where $k=2m\pi /N$, $m=0,1,2,...,N-1$, and $\alpha _{k},$\ $\beta _{k}$ are
fermionic operators in $k$ space defined by%
\begin{equation}
\left\{
\begin{array}{cc}
\alpha _{k}=\frac{1}{\sqrt{N}}\sum\limits_{j}e^{-ik\left( j+1\right)
/2}c_{j}, & \text{even }j \\
\beta _{k}=\frac{1}{\sqrt{N}}\sum\limits_{j}e^{-ik\left( j+1\right)
/2}c_{j},, & \text{odd }j%
\end{array}%
\right. .
\end{equation}%
This transformation block diagonalizes the Hamiltonian due to its
translational symmetry, i.e.,%
\begin{equation}
H=\sum_{k\geq 0}H_{k},
\end{equation}%
satisfying $\left[ H_{k},H_{k^{\prime }}\right] =\delta _{kk^{\prime }}$,
where%
\begin{eqnarray}
&&H_{k}=-2\cos \frac{k}{2}\left( \alpha _{k}^{\dagger }\beta _{k}+\alpha
_{-k}^{\dagger }\beta _{-k}\right)  \notag \\
&&+i2\sin \frac{k}{2}\left( \alpha _{-k}^{\dagger }\beta _{k}^{\dagger
}-\alpha _{k}^{\dagger }\beta _{-k}^{\dagger }\right) +\mathrm{h.c.}  \notag
\\
&&+2g_{2}\left( \alpha _{k}^{\dagger }\alpha _{k}-\alpha _{-k}\alpha
_{-k}^{\dagger }\right) +2g_{1}\left( \beta _{k}^{\dag }\beta _{k}-\beta
_{-k}\beta _{-k}^{\dag }\right) ,
\end{eqnarray}%
In order to diagonalize\ $H_{k}$, we rewrite $H_{k}$ in the basis%
\begin{equation}
\psi _{k}^{\dag }=\left(
\begin{array}{cccc}
\alpha _{k}^{\dag } & \beta _{k}^{\dagger } & \alpha _{-k} & \beta _{-k}%
\end{array}%
\right) ,\psi _{k}=\left(
\begin{array}{c}
\alpha _{k} \\
\beta _{k} \\
\alpha _{-k}^{\dag } \\
\beta _{-k}^{\dagger }%
\end{array}%
\right) ,
\end{equation}%
in the Nambu representation%
\begin{equation}
H_{k}=\psi _{k}^{\dag }h_{k}\psi _{k}.
\end{equation}%
Here $h_{k}$ is a $4\times 4$\ matrix

\begin{equation}
\frac{h_{k}}{2}=\left(
\begin{array}{cccc}
g_{2} & -\cos \frac{k}{2} & 0 & -i\sin \frac{k}{2} \\
-\cos \frac{k}{2} & g_{1} & -i\sin \frac{k}{2} & 0 \\
0 & i\sin \frac{k}{2} & -g_{2} & \cos \frac{k}{2} \\
i\sin \frac{k}{2} & 0 & \cos \frac{k}{2} & -g_{1}%
\end{array}%
\right) ,  \label{h_k}
\end{equation}%
the diagonalization of which leads to the solution of the original
Hamiltonian in both Hermitian and non-Hermitian versions.

\subsection{Real field}

In this case, matrix $h_{k}$\ is Hermitian and can be diagonalized
directively. The eigenvectors $\Psi _{\rho \sigma }$ and eigenvalues $%
\varepsilon _{\rho \sigma }$, obeying the equation $h_{k}\Psi _{\rho \sigma
}=\varepsilon _{\rho \sigma }\Psi _{\rho \sigma }$, are%
\begin{eqnarray}
\Psi _{\rho \sigma } &=&\frac{1}{\sqrt{\Omega _{\rho \sigma }}}\left(
\begin{array}{c}
\eta _{\rho }^{\sigma } \\
\xi _{\rho }^{\sigma } \\
-2g_{1}\sin k \\
\Lambda _{\rho }^{\sigma }%
\end{array}%
\right) ,  \label{eigenvector} \\
\varepsilon _{\rho \sigma } &=&-\rho \epsilon _{\sigma }.  \label{spectrum}
\end{eqnarray}%
Here the $k$-dependent factors $\left\{ \Lambda _{\rho }^{\sigma },\eta
_{\rho }^{\sigma },\xi _{\rho }^{\sigma },\epsilon _{\sigma },A\right\} $\
are explicitly expressed as
\begin{eqnarray}
\Lambda _{\rho }^{\sigma } &=&\sin \frac{k}{2}[\left( g_{1}-g_{2}\right)
^{2}+\sigma A+\rho \epsilon _{\sigma }\left( g_{1}-g_{2}\right) ], \\
\eta _{\rho }^{\sigma } &=&-i[2g_{1}\cos k+2g_{2}  \notag \\
&&+\frac{\rho \epsilon _{\sigma }-2g_{2}}{2}\left(
g_{1}^{2}-g_{2}^{2}-\sigma A\right) ], \\
\xi _{\rho }^{\sigma } &=&i\cos \frac{k}{2}[\left( g_{1}+g_{2}\right)
^{2}+\sigma A-\rho \epsilon _{\sigma }\left( g_{1}+g_{2}\right) ],
\end{eqnarray}%
and%
\begin{eqnarray}
\Omega _{\rho \sigma } &=&\left( \eta _{\rho }^{\sigma }\right) ^{2}+\left(
\xi _{\rho }^{\sigma }\right) ^{2}+4g_{1}^{2}\sin ^{2}k+\left( \Lambda
_{\rho }^{\sigma }\right) ^{2} \\
\epsilon _{\sigma } &=&\sqrt{2}\sqrt{g_{1}^{2}+g_{2}^{2}+\sigma A+2}, \\
A &=&[\left( g_{1}^{2}-g_{2}^{2}\right) ^{2}+4g_{1}^{2}+8g_{1}g_{2}\cos
k+4g_{2}^{2}]^{1/2},
\end{eqnarray}%
with $\sigma ,\rho =\pm $. Based on this result, the Hamiltonian can be
diagonalized in the form

\begin{equation}
H=\sum_{k\geq 0,\rho ,\sigma }\varepsilon _{\rho \sigma }\left( \gamma
_{\rho \sigma }^{k}\right) ^{\dag }\gamma _{\rho \sigma }^{k},  \label{H_d1}
\end{equation}%
where the fermion operator is defined as

\begin{equation}
\gamma _{\rho \sigma }^{k}=\left( \Psi _{\rho \sigma }\right) ^{T}\psi _{k}.
\label{gamma}
\end{equation}%
The ground state is%
\begin{equation}
\left\vert \mathrm{G}\right\rangle =\prod_{k\geq 0,\sigma =\pm }\left(
\gamma _{+\sigma }^{k}\right) ^{\dag }\left\vert \mathrm{Vac}\right\rangle ,
\end{equation}%
with the groundstate energy density%
\begin{equation}
\frac{E_{\mathrm{g}}}{2N}=\frac{1}{2N}\sum_{k\geq 0,\sigma =\pm }\varepsilon
_{+\sigma }(k),
\end{equation}%
where $\left\vert \mathrm{Vac}\right\rangle $\ is the vacuum state of
operator $\gamma _{\rho \sigma }^{k}$. The phase diagram can be identified
by the behavior of $E_{\mathrm{g}}/\left( 2N\right) $. We note that $E_{%
\mathrm{g}}$\ is the summation of pair $\sum_{\sigma }\varepsilon _{+\sigma
}=\varepsilon _{++}+\varepsilon _{+-}$ for each $k$:%
\begin{eqnarray}
&&\sum_{\sigma }\varepsilon _{+\sigma }=-2\{g_{1}^{2}+g_{2}^{2}+2  \notag \\
&&+2[\left( g_{1}g_{2}-\cos k\right) ^{2}+\sin ^{2}k]^{1/2}\}^{1/2}.
\end{eqnarray}%
Obviously, around points $k=0,\pi $, the term $\left\vert g_{1}g_{2}\pm
1\right\vert $ leads to the discontinuity of the derivative of groundstate
density. Then the phase boundary is the line%
\begin{equation}
\left\vert g_{1}g_{2}\right\vert =1,
\end{equation}%
which accords with the result for the ordinary transverse Ising model when
we take $g_{1}=g_{2}$. The phase diagram is illustrated in Fig. \ref{fig1}%
(a).

\subsection{Complex field}

In this case, matrix $h_{k}$\ is non-Hermitian and can also be diagonalized
directively. The eigenvectors and eigenvalues are still in the form of Eqs. (%
\ref{eigenvector}) and (\ref{spectrum}). The factors $\left\{ \Lambda _{\rho
}^{\sigma },\eta _{\rho }^{\sigma },\xi _{\rho }^{\sigma },\epsilon _{\sigma
},A\right\} $ can be obtained by taking $g_{_{1}}=\eta -i\xi $\ and $%
g_{2}=\eta +i\xi $, while the normalization factor $\Omega _{\rho \sigma }$
should be redefined based on the eigenvector of matrix $h_{k}^{\dag }$.
Similarly, the eigenvectors and eigenvalues of $h_{k}^{\dagger }$\ are still
in the form of Eqs. (\ref{eigenvector}) and (\ref{spectrum}), with the
factors $\left\{ \Lambda _{\rho }^{\sigma },\eta _{\rho }^{\sigma },\xi
_{\rho }^{\sigma },\epsilon _{\sigma },A\right\} $ by taking $g_{_{1}}=\eta
+i\xi $\ and $g_{2}=\eta -i\xi $.

The non-Hermitian Hamiltonian can be diagonalized in the form

\begin{equation}
H=\sum_{k\geq 0,\rho ,\sigma }\varepsilon _{\rho \sigma }\overline{\zeta }%
_{\rho \sigma }^{k}\zeta _{\rho \sigma }^{k},  \label{H_d2}
\end{equation}%
where the fermion operators $\overline{\zeta }_{\rho \sigma }^{k}$\ and $%
\zeta _{\rho \sigma }^{k}$ are defined by%
\begin{eqnarray}
\zeta _{\rho \sigma }^{k} &=&\gamma _{\rho \sigma }^{k}\left(
g_{_{1}}\rightarrow \eta -i\xi ,g_{2}\rightarrow \eta +i\xi \right) , \\
\overline{\zeta }_{\rho \sigma }^{k} &=&\left( \gamma _{\rho \sigma
}^{k}\right) ^{\dag }\left( g_{_{1}}\rightarrow \eta +i\xi ,g_{2}\rightarrow
\eta -i\xi \right) .
\end{eqnarray}%
Note that $\overline{\zeta }_{\rho \sigma }^{k}\neq \left( \zeta _{\rho
\sigma }^{k}\right) ^{\dag }$, but
\begin{equation}
\{\zeta _{\rho ^{\prime }\sigma ^{\prime }}^{k\prime },\overline{\zeta }%
_{\rho \sigma }^{k}\}=\delta _{kk\prime }\delta _{\rho \rho ^{\prime
}}\delta _{\sigma \sigma ^{\prime }}.
\end{equation}%
The ground states of $H$ and $H^{\dag }$ can be constructed as%
\begin{equation}
\left\vert \mathrm{G}\right\rangle =\prod_{k\geq 0,\sigma =\pm }\overline{%
\zeta }_{+\sigma }^{k}\left\vert \mathrm{Vac}\right\rangle ,
\end{equation}%
and%
\begin{equation}
\left\langle \overline{\mathrm{G}}\right\vert =\left\langle \overline{%
\mathrm{Vac}}\right\vert \prod_{k\geq 0,\sigma =\pm }\zeta _{+\sigma }^{k},
\end{equation}%
respectively. Here, vacuum states are defined by $\zeta _{\rho \sigma
}^{k}\left\vert \mathrm{Vac}\right\rangle =0$ and $\left\langle \overline{%
\mathrm{Vac}}\right\vert \left( \overline{\zeta }_{\rho \sigma }^{k}\right)
^{\dag }=0$.

Accordingly, the phase diagram can be identified by the behavior of the term%
\begin{eqnarray}
&&\sum_{\sigma }\varepsilon _{+\sigma }=-2\{\eta ^{2}-\xi ^{2}+1  \notag \\
&&+2[\left( \eta ^{2}+\xi ^{2}-\cos k\right) ^{2}+\sin ^{2}k]^{1/2}\}^{1/2},
\end{eqnarray}%
when $k\rightarrow 0$%
\begin{equation}
\varepsilon _{++}+\varepsilon _{+-}=-2[\left( \eta ^{2}-\xi ^{2}+1\right)
+2\left\vert \eta ^{2}+\xi ^{2}-1\right\vert ]^{1/2}.
\end{equation}%
The term $\left\vert \eta ^{2}+\xi ^{2}-1\right\vert $\ leads to the
discontinuity of the derivative of groundstate density. Then the phase
boundary are the lines%
\begin{equation}
\eta ^{2}+\xi ^{2}=1
\end{equation}%
separating paramagnetic and ferromagnetic phases \cite{LC1 PRA}. The phase
diagram is illustrated in Fig. \ref{fig1}(b). The aim of this paper is
trying to retrieve the quantum phase diagrams by some geometric quantities
of the models.

\begin{figure*}[tbp]
\centering
\includegraphics[ bb=31 243 333 491, width=0.16\textwidth, clip]{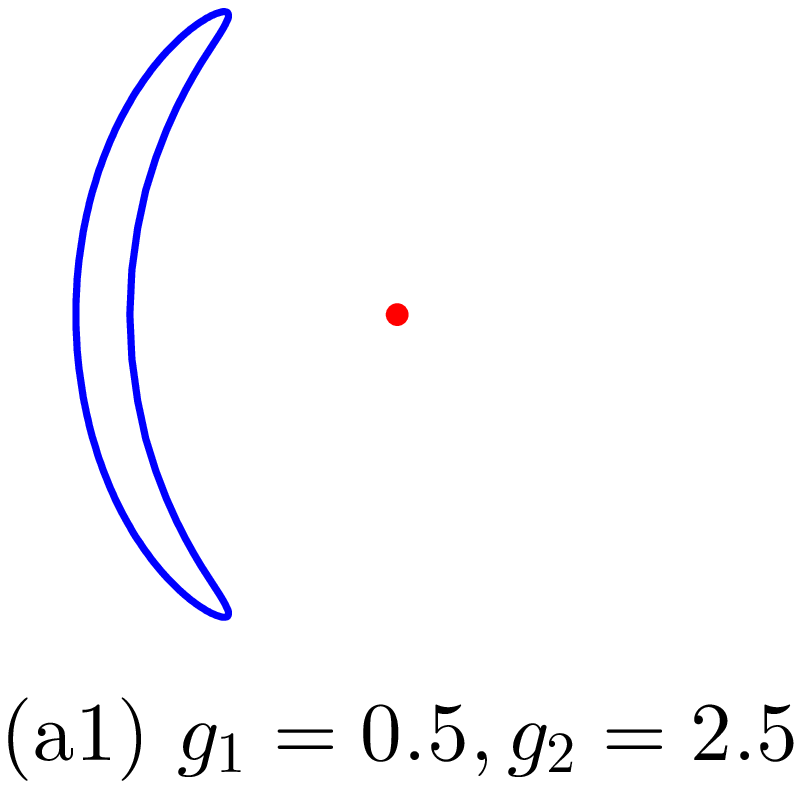} %
\includegraphics[ bb=31 243 333 491, width=0.16\textwidth, clip]{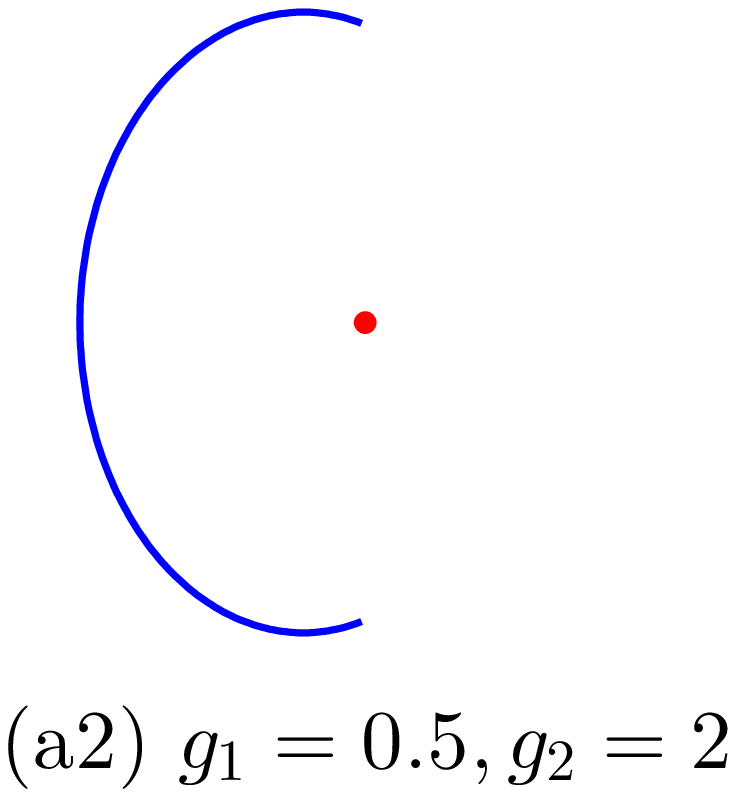} %
\includegraphics[ bb=31 243 333 491, width=0.16\textwidth, clip]{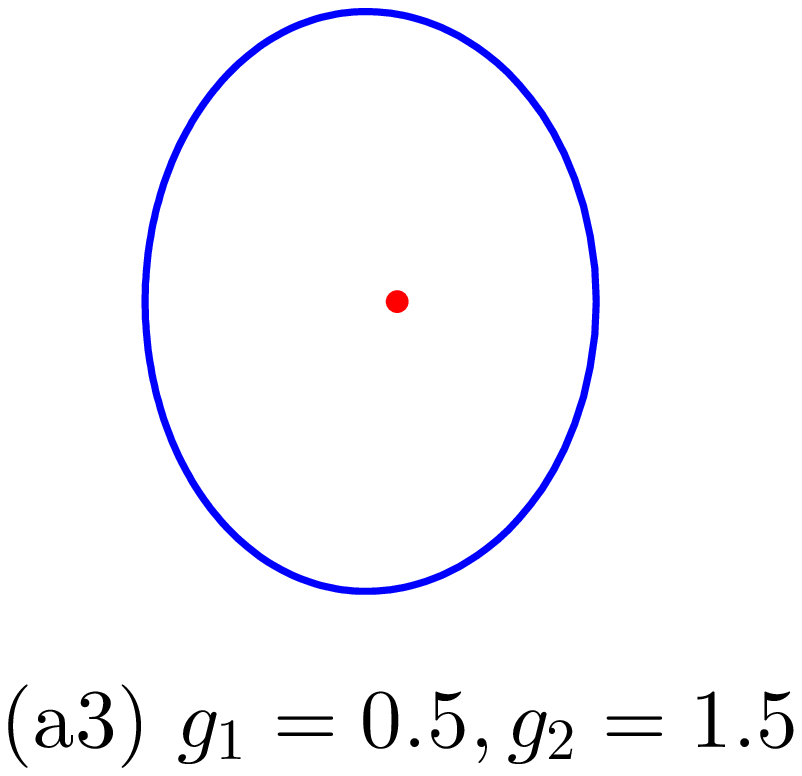} %
\includegraphics[ bb=31 243 333 491, width=0.16\textwidth, clip]{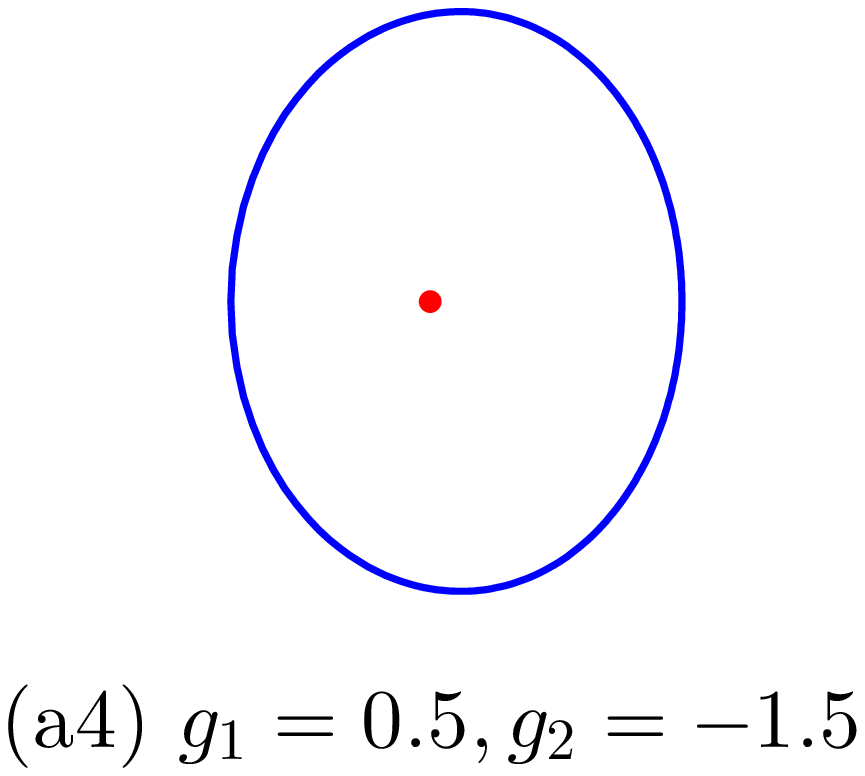} %
\includegraphics[ bb=31 243 333 491, width=0.16\textwidth, clip]{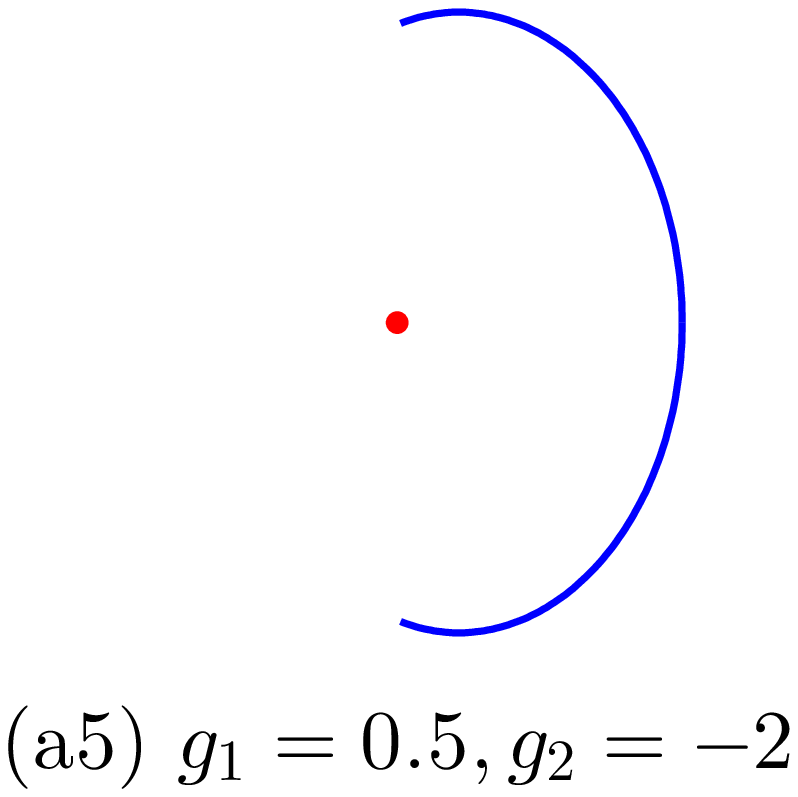} %
\includegraphics[ bb=31 243 333 491, width=0.16\textwidth, clip]{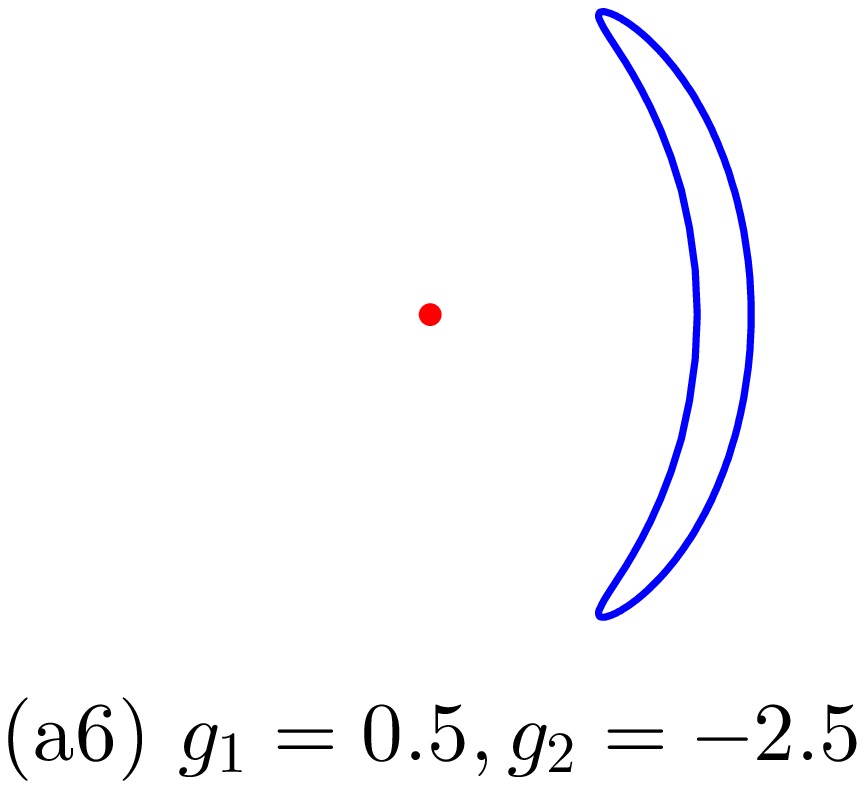} %
\includegraphics[ bb=31 243 333 491, width=0.16\textwidth, clip]{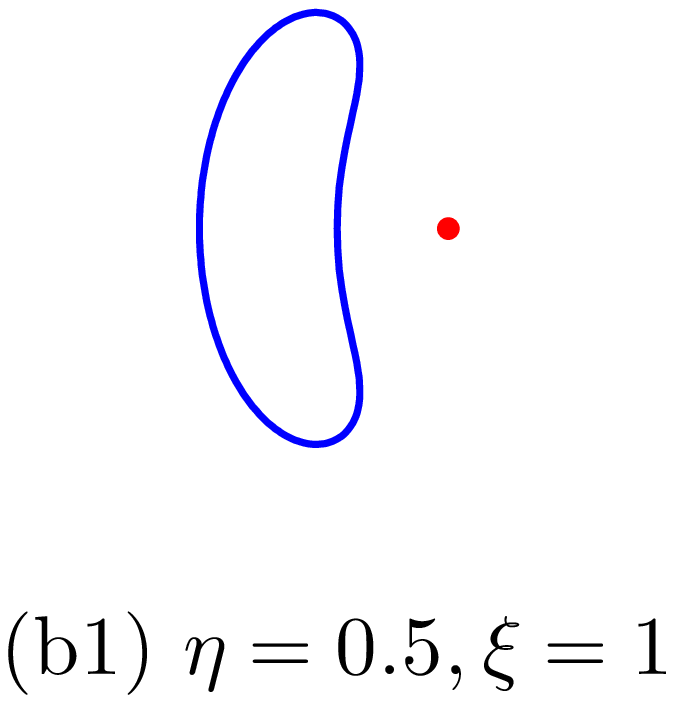} %
\includegraphics[ bb=31 243 333 491, width=0.16\textwidth, clip]{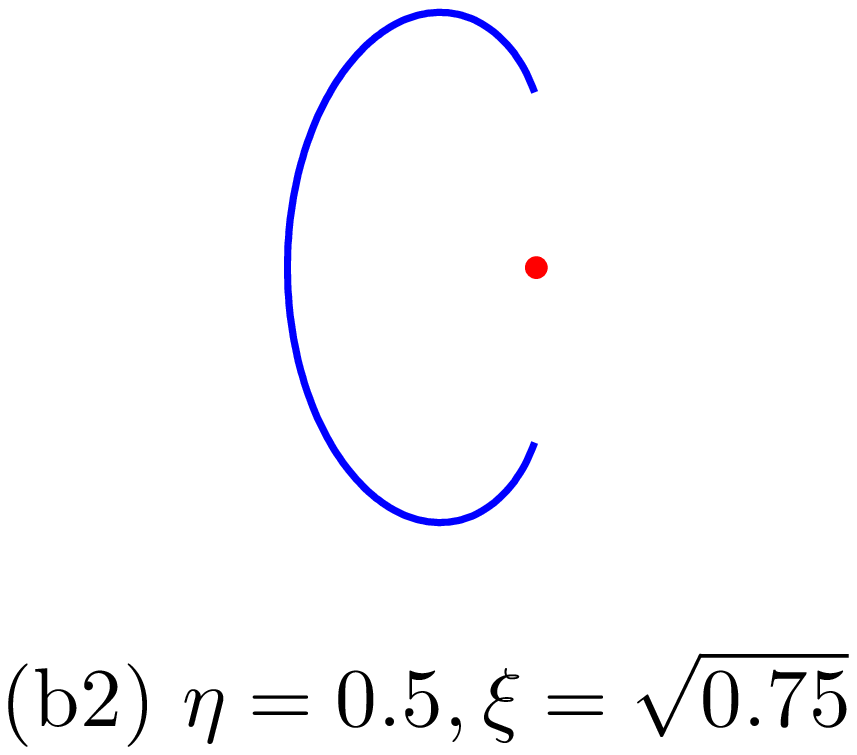} %
\includegraphics[ bb=31 243 333 491, width=0.16\textwidth, clip]{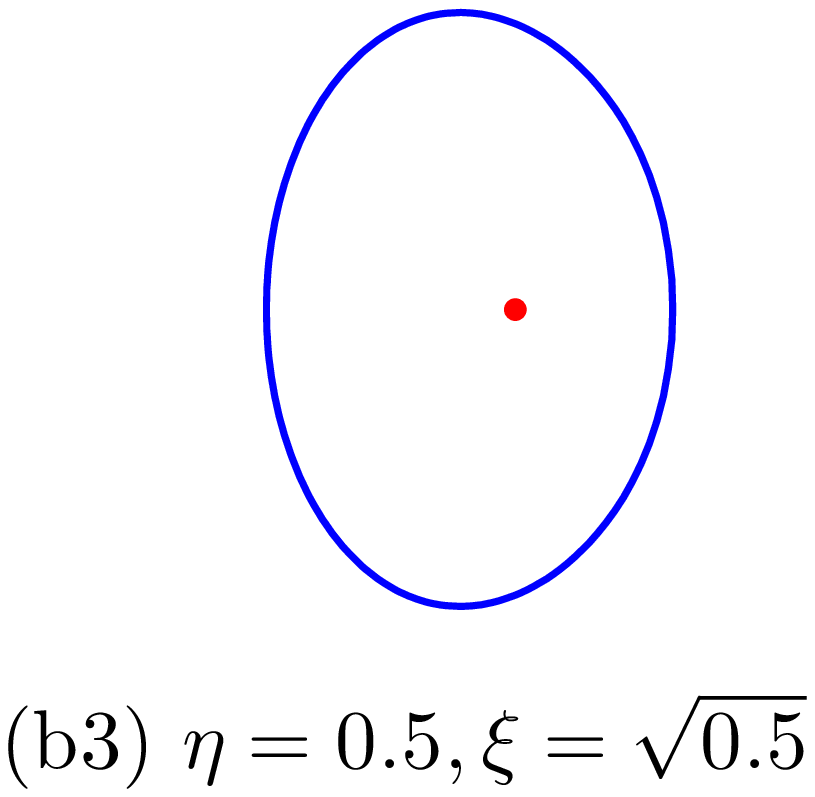} %
\includegraphics[ bb=31 243 333 491, width=0.16\textwidth, clip]{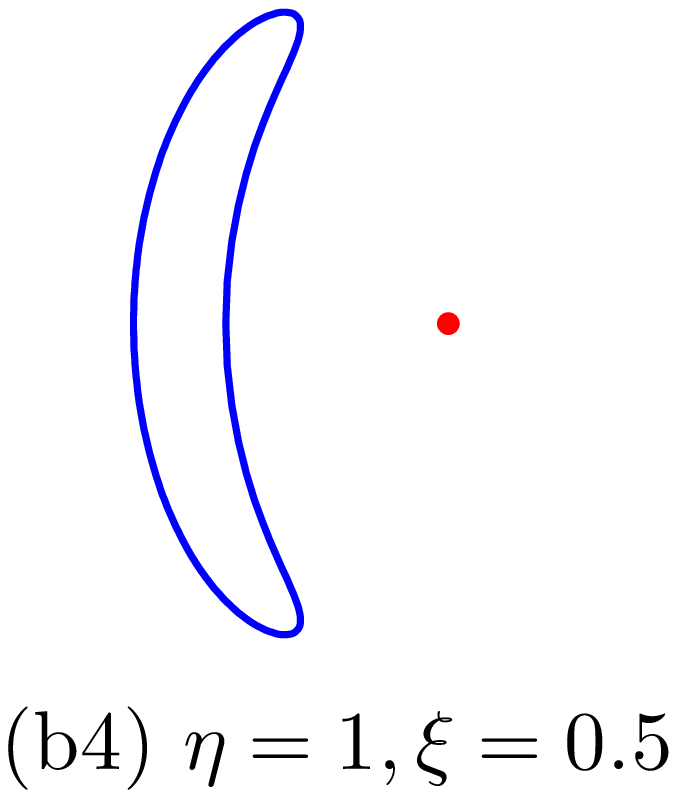} %
\includegraphics[ bb=31 243 333 491, width=0.16\textwidth, clip]{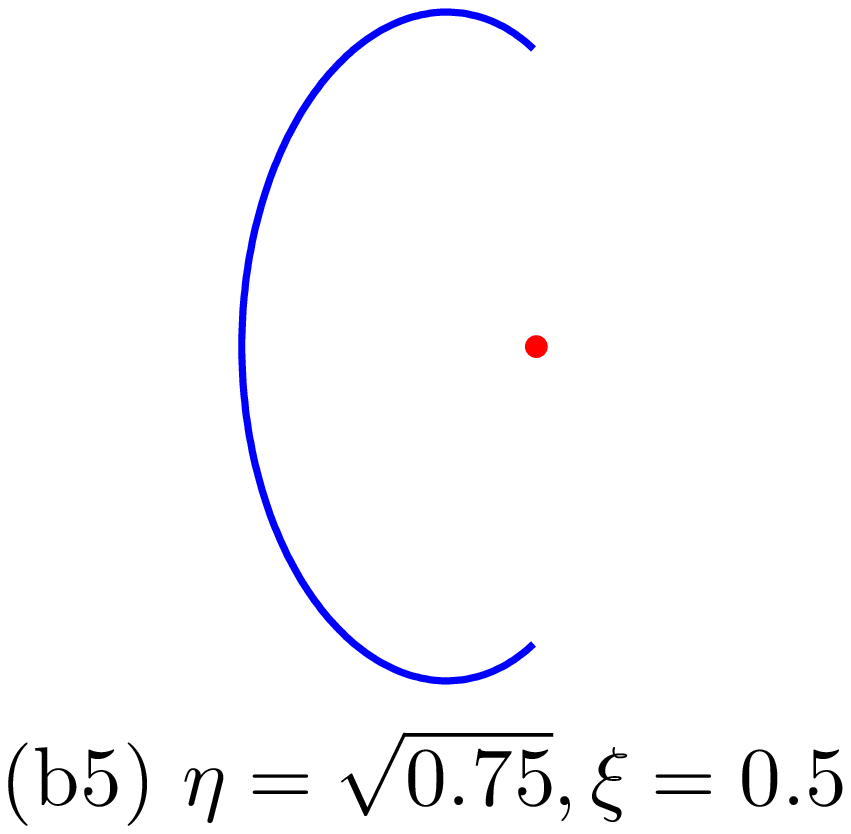} %
\includegraphics[ bb=31 243 333 491, width=0.16\textwidth, clip]{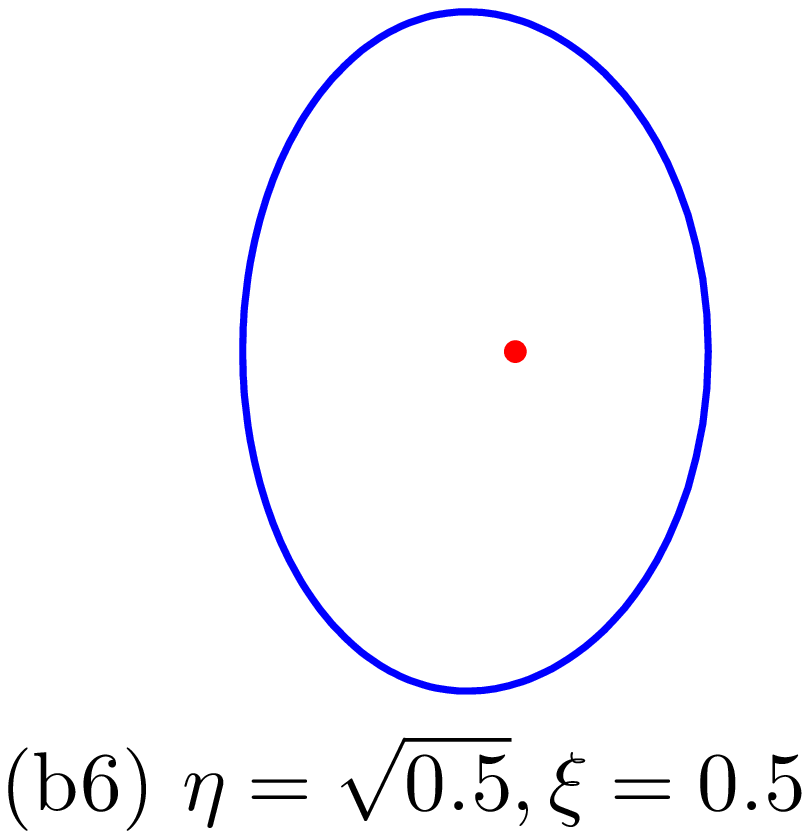}
\caption{(color online) Several typical graphs obtained from parameter
equations in Eq. \protect\ref{loop}, which characterize the quantum phases
for Hermitian and non-Hermitian Ising models. The red dot indicates the
origin of the $xy$ plane.}
\label{fig2}
\end{figure*}

\section{Chern numbers}

\label{Chern numbers}

In the previous work \cite{LC1 PRA}, it was found that the Berry phase can
be utilized to identify the phase diagram of the non-Hermitian Hamiltonian.
We now investigate the connection between quantum phases and other geometric
quantity, the Chern number. We start our analysis from the Bloch Hamiltonian
$H_{k}$. We will give the derivations in parallel steps for both Hermitian
and non-Hermitian Hamiltonians. In both cases, one can rewrite the
Hamiltonian in the form%
\begin{equation}
H_{k}=\vec{B}_{k}\cdot \vec{s}_{k},
\end{equation}%
where $\vec{B}_{k}=(0,0,2(\varepsilon _{++}+\varepsilon _{+-}))$\ and pseudo
spin operators are defined as%
\begin{eqnarray}
s_{k}^{-} &=&\left( s_{k}^{+}\right) ^{\dag }=\gamma _{++}^{k}\gamma
_{+-}^{k},  \notag \\
s_{k}^{z} &=&\frac{1}{2}[\sum_{\sigma =\pm }\left( \gamma _{+\sigma
}^{k}\right) ^{\dag }\gamma _{+\sigma }^{k}-1],
\end{eqnarray}%
for real field and%
\begin{eqnarray}
s_{k}^{-} &=&\zeta _{++}^{k}\zeta _{+-}^{k},s_{k}^{+}=\overline{\zeta }%
_{+-}^{k}\overline{\zeta }_{++}^{k},  \notag \\
s_{k}^{z} &=&\frac{1}{2}[\sum_{\sigma =\pm }\overline{\zeta }_{+\sigma
}^{k}\zeta _{+\sigma }^{k}-1],
\end{eqnarray}%
for complex field, satisfying the commutation relations of Lie algebra%
\begin{equation}
\left[ s_{k}^{z},s_{k^{\prime }}^{\pm }\right] =\pm \delta _{kk^{\prime
}}s_{k^{\prime }}^{\pm },\text{ }\left[ s_{k}^{+},s_{k^{\prime }}^{-}\right]
=2\delta _{kk^{\prime }}s_{k^{\prime }}^{z}.
\end{equation}%
In order to capture the topological features of the ground states, we map
the Hamiltonian $H_{k}$\ to $H_{k}(\theta ,\varphi )$\textbf{.} It has the
form%
\begin{equation}
H_{k}(\theta ,\varphi )=\vec{B}_{k}(\theta ,\varphi )\cdot \vec{s}_{k},
\end{equation}%
where the field is%
\begin{equation}
\vec{B}_{k}(\theta ,\varphi )=(\left\vert \vec{B}_{k}\right\vert \sin
\varphi \cos \theta ,\left\vert \vec{B}_{k}\right\vert \sin \varphi \sin
\theta ,\cos \varphi ).
\end{equation}%
Here $\theta $ is taken as%
\begin{eqnarray}
&&\tan \theta _{k}=\frac{\sin k}{\cos k-g_{1}g_{2}},\text{ }  \notag \\
&&\text{and }\frac{\sin k}{\cos k-(\eta ^{2}+\xi ^{2})},  \label{theta}
\end{eqnarray}%
for real and complex fields, respectively. We note that when $\varphi =\pi
/2 $,\ $H_{k}(\theta ,\varphi )$\ has the same spectrum with the original
Hamiltonian\textbf{\ }$H_{k}$\textbf{.}

The eigenstates of $H_{k}(\theta ,\varphi )$ in the even particle-number
invariant subspace are

\begin{eqnarray}
\left\vert u_{k}^{\pm }\right\rangle &=&\frac{1}{\sqrt{\Omega _{k}^{\pm }}}%
[(\cos \varphi +\varepsilon _{k}^{\pm })(\gamma _{++}^{k}\gamma
_{+-}^{k})^{\dag }  \notag \\
&&+\left\vert \vec{B}_{k}\right\vert ^{2}e^{i\theta }\sin \varphi
]\left\vert \mathrm{Vac}\right\rangle , \\
&&\text{and }\frac{1}{\sqrt{\Omega _{k}^{\pm }}}[(\cos \varphi +\varepsilon
_{k}^{\pm })\overline{\zeta }_{+-}^{k}\overline{\zeta }_{++}^{k}  \notag \\
&&+\left\vert \vec{B}_{k}\right\vert ^{2}e^{i\theta }\sin \varphi
]\left\vert \mathrm{Vac}\right\rangle ,
\end{eqnarray}%
with $\Omega _{k}^{\pm }=2\left\vert \varepsilon _{k}^{\pm }\right\vert
\left( \left\vert \varepsilon _{k}^{\pm }\right\vert \pm \cos \varphi
\right) $, for real and complex fields, respectively. The corresponding
eigen energy is%
\begin{equation}
\varepsilon _{k}^{\pm }=\pm \sqrt{\cos ^{2}\varphi +\left\vert \vec{B}%
_{k}\right\vert ^{2}\sin ^{2}\varphi }.
\end{equation}%
We are interested in the ground state, then considering the lower energy
level. The Berry connection for $\left\vert u_{k}^{-}\right\rangle $\ is
given by%
\begin{equation}
A_{k}=i\left\langle u_{k}^{-}\right\vert \partial _{k}\left\vert
u_{k}^{-}\right\rangle ,A_{\varphi }=i\left\langle u_{k}^{-}\right\vert
\partial _{\varphi }\left\vert u_{k}^{-}\right\rangle ,
\end{equation}%
for real field and%
\begin{equation}
A_{k}=i\left\langle \bar{u}_{k}^{-}\right\vert \partial _{k}\left\vert
u_{k}^{-}\right\rangle ,A_{\varphi }=i\left\langle \bar{u}%
_{k}^{-}\right\vert \partial _{\varphi }\left\vert u_{k}^{-}\right\rangle ,
\end{equation}%
for complex field, where $\left\langle \bar{u}_{k}^{-}\right\vert $\ is the
biorthonormal conjugation of $\left\vert u_{k}^{-}\right\rangle $. We would
like to point out that the Berry connection here is defined in the framework
of biorthonormal inner product for the non-Hermitian system. A
straightforward derivation shows that, in both cases, we have%
\begin{eqnarray}
A_{k} &=&-\frac{1}{\Omega _{k}^{-}}\left\vert \vec{B}_{k}\right\vert
^{2}\sin ^{2}\varphi \frac{\partial \theta }{\partial k}, \\
A_{\varphi } &=&0.
\end{eqnarray}%
and the Berry curvature is%
\begin{eqnarray}
\Omega _{k\varphi } &=&\partial _{k}A_{\varphi }-\partial _{\varphi }A_{k}
\notag \\
&=&\frac{1}{2(\varepsilon _{k}^{-})^{3}}\left\vert \vec{B}_{k}\right\vert
^{2}\sin \varphi \frac{\partial \theta }{\partial k}.
\end{eqnarray}%
For the cases with $\left\vert g_{1}g_{2}\right\vert \neq 1$\ and $\eta
^{2}+\xi ^{2}\neq 1$, the corresponding Chern number is%
\begin{eqnarray}
c &=&\frac{1}{2\pi }\int_{0}^{\pi }\text{d}\varphi \int_{0}^{2\pi }\text{d}%
k\Omega _{k\varphi }  \notag \\
&=&-\frac{1}{2\pi }\left[ \theta \left( 2\pi \right) -\theta \left( 0\right) %
\right] ,  \label{chern_number}
\end{eqnarray}%
which only depends on the function of $\theta \left( k\right) $\ defined in
Eq. (\ref{theta}). For the cases with $\left\vert g_{1}g_{2}\right\vert =1$
and $\eta ^{2}+\xi ^{2}=1$, a straightforward derivation shows $c=\frac{1}{2}
$. In summary, it is easy to check that the Chern number is

\begin{equation}
c=\left\{
\begin{array}{cc}
1, & \left\vert g_{1}g_{2}\right\vert <1 \\
\frac{1}{2} & \left\vert g_{1}g_{2}\right\vert =1 \\
0, & \left\vert g_{1}g_{2}\right\vert >1%
\end{array}%
\right. ,
\end{equation}%
for the Hermitian system and%
\begin{equation}
c=\left\{
\begin{array}{cc}
1, & \eta ^{2}+\xi ^{2}<1 \\
\frac{1}{2} & \eta ^{2}+\xi ^{2}=1 \\
0, & \eta ^{2}+\xi ^{2}>1%
\end{array}%
\right. ,
\end{equation}%
for the non-Hermitian system, which accords with the phase diagram. We get
the conclusion that the topological quantity, the Chern number can be used
to identify the quantum phases. It indicates that the connection between QPT
and topological quantity in the Ising model can be extended to more general
systems.

This conclusion can be understood from the viewpoint of geometry. We
consider the case with $\varphi =\pi /2$, the field reduces to a
two-dimensional field. In this plane, the ground state corresponds to a loop
depicted by the parameter equations%
\begin{equation}
x=\left\vert \vec{B}_{k}\right\vert \cos \theta ,y=\left\vert \vec{B}%
_{k}\right\vert \sin \theta .  \label{loop}
\end{equation}%
The winding number of a closed curve in the auxiliary $xy$-plane around the
origin is defined as

\begin{equation}
\nu =\frac{1}{2\pi }\int\nolimits_{c}\frac{1}{\left\vert \vec{B}%
_{k}\right\vert ^{2}}\left( x\mathrm{d}y-y\mathrm{d}x\right) ,
\label{winding number}
\end{equation}%
which has been shown to equal to the Chern number \cite{ZG Arxiv}. To
illustrate this point, we plot the graphs with several typical values of $%
g_{1}g_{2}$\ and $\eta ^{2}+\xi ^{2}$\ in Fig. \ref{fig2}. It clearly shows
the connections between the quantum phase diagram and the geometry of the
graphs.

\section{Summary}

\label{Summary}

We have studied the QPTs in an one-dimensional transverse field Ising model
with additional periodically modulated real and complex fields from an
alternative view. The equivalent Hamiltonians in the $k$ space allow us to
characterize the change of the ground states on passing the phase
transition. Based on the exact solution, we have found that the phase
boundaries can be identified by the Chern numbers defined in the context of
Dirac and biorthonormal inner products, respectively. A notable feature
which emerges is that the indicator of the phase transition is not only a
local order parameter but also can be the Chern number, which is a key
topological quantity. Furthermore, it is available for the non-Hermitian
system. An interesting prediction is the extension of such methods to a
wider class of quantum spin systems which is consisted of multi-sublattices
with the periodic modulation on the coupling strength and field.

\acknowledgments We acknowledge the support of the National Basic Research
Program (973 Program) of China under Grant No. 2012CB921900 and CNSF (Grant
No. 11374163).\newline

\end{document}